\documentclass[12pt,preprint]{aastex}
\usepackage{epsfig}
\usepackage{natbib}
\usepackage{graphicx}
\usepackage{slashbox}
\usepackage{multirow}
\usepackage{lscape}
\usepackage{mathrsfs,amssymb}
\usepackage{amsmath}
\usepackage{subfigure}
\usepackage{amssymb}
\usepackage{multirow}
\usepackage{tabularx}
\usepackage{rotating}
\usepackage{amsmath}

\newcommand       \cm           {\,{\rm cm}}

\newcommand       \K            {\,{\rm K}}

\newcommand       \simgt        {\gtrsim}

\newcommand       \mum          {\,{\rm \mu m}}

\newcommand       \simali       {\sim\,}

\newcommand       \Iaro         {I_{3.3}}
\newcommand       \Iali         {I_{3.4}}
\newcommand       \Iratio         {I_{3.4}/I_{3.3}}
\newcommand       \Aratio        {A_{3.4}/A_{3.3}}
\newcommand       \Aali           {A_{3.4}}
\newcommand       \Aaro          {A_{3.3}}
\newcommand       \km        {\,{\rm km}}
\newcommand       \mol       {\,{\rm mol}}
\newcommand       \cals       {\,{\rm cal}}
\newcommand       \kcal       {\,{\rm kcal}}

\newcommand       \Etot      {E_{\rm tot}}

\newcommand       \tauaro    {\tau_{\rm arom}}
\newcommand       \tauali    {\tau_{\rm aliph}}

\countdef\decade=200
\advance\decade by \year
\countdef\hours=201
\advance\hours by \time
\divide\hours by 60
\countdef\mins=202
\advance\mins by \hours
\multiply\mins by 60
\multiply\hours by 100
\countdef\miltime=203
\advance\miltime by \hours
\advance\miltime by \time
\advance\miltime by -\mins
\def\today{\number\decade.\number\month.\number\day.\number\miltime}

\shorttitle{C--H Stretches of PAHs with Aliphatic Substituents}
\title{
\vspace*{-2.0em}
{\normalsize\rm Accepted for publication in {\it The Astrophysical Journal}}\\
\vspace*{1.0em}
The C--H Stretching Features at 3.2--3.5$\mum$
of Polycyclic Aromatic Hydrocarbons with Aliphatic Sidegroups
\\{\small DRAFT: \today ~~}
}
\author{X.J.~Yang\altaffilmark{1,2},
        Aigen Li\altaffilmark{2},
        R.~Glaser\altaffilmark{3},
        and J.X.~Zhong\altaffilmark{1}}
\altaffiltext{1}{Department of Physics,
                      Xiangtan University,
                      411105 Xiangtan, Hunan Province, China;
                      {\sf xjyang@xtu.edu.cn, jxzhong@xtu.edu.cn}}
\altaffiltext{2} {Department of Physics and Astronomy,
                  University of Missouri,
                  Columbia, MO 65211, USA;
                  {\sf lia@missouri.edu}}
\altaffiltext{3} {Department of Chemistry,
                  University of Missouri,
                  Columbia, MO 65211, USA;
                  {\sf glaserr@missouri.edu}}

\begin{document}

\begin{abstract}
The so-called ``unidentified'' infrared emission (UIE) features
at 3.3, 6.2, 7.7, 8.6, and 11.3$\mum$ are ubiquitously
seen in a wide variety of astrophysical regions.
The UIE features are characteristic of
the stretching and bending vibrations of aromatic
hydrocarbon materials,
e.g., polycyclic aromatic hydrocarbon (PAH) molecules.
The 3.3$\mum$ aromatic C--H stretching feature
is often accompanied by a weaker feature at 3.4$\mum$.
The latter is generally thought to result from
the C--H stretch of aliphatic groups attached
to the aromatic systems.
The ratio of the observed intensity of
the 3.3$\mum$ aromatic C--H feature 
to that of the 3.4$\mum$ aliphatic C--H feature 
allows one to estimate the aliphatic fraction
of the UIE carriers,
provided that the intrinsic oscillator strengths
of the 3.3$\mum$ aromatic
C--H stretch ($\Aaro$) and the 3.4$\mum$ aliphatic
C--H stretch ($\Aali$) are known.
%
While previous studies on the aliphatic fraction
of the UIE carriers were mostly based on
the $\Aratio$ ratios
derived from the mono-methyl derivatives of
small PAH molecules,
in this work we employ density functional theory
to compute the infrared vibrational spectra of
PAH molecules with a wide range of sidegroups
including ethyl, propyl, butyl,
and several unsaturated alkyl chains,
as well as all the isomers of dimethyl-substituted pyrene.
We find that, except PAHs with unsaturated alkyl chains,
the corresponding $\Aratio$ ratios are close
to that of mono-methyl PAHs.
This confirms the predominantly-aromatic nature
of the UIE carriers previously inferred from
the $\Aratio$ ratio derived from mono-methyl PAHs.
\end{abstract}
\keywords {dust, extinction --- ISM: lines and bands
           --- ISM: molecules}

\section{Introduction\label{sec:intro}}
Over four decades ago,
an important chapter of modern astrochemistry 
was opened by Gillett et al.\ (1973) 
who obtained from ground-based observations 
the $\simali$8--13$\mum$ spectra of two planetary nebulae 
(NGC\,7027 and BD$+$30$^{\rm o}$3639) and reported
the discovery of a broad emission feature at 11.3$\mum$.
Two additional emission features 
peaking at 8.6 and 12.7$\mum$
were also prominent in their spectra.
Since then other features have been observed,
including the 6.2 and 7.7$\mum$ bands 
first discovered in NGC\,7027
with airborne observations 
made with the {\it Kuiper Airborne Observatory} 
(Russell et al.\ 1977)
and the 3.3$\mum$ band
first discovered also in NGC\,7027
through ground-based observations
(Merrill et al.\ 1975). 
It had already became clear in the 1970s that
these bands were part of a family of 
infrared (IR) emission features.   
Now, this family of emission features 
at 3.3, 6.2, 7.7, 8.6, 11.3, and 12.7$\mum$ 
are ubiquitously seen in almost all celestial objects 
with associated gas and dust, including 
protoplanetary nebulae (PPNe),
planetary nebulae (PNe), 
protoplanetary disks around young stars,
reflection nebulae, HII regions,
the Galactic IR cirrus,
and external galaxies (see Tielens 2008).
Collectively known as 
the  ``unidentified infrared emission'' (UIE) bands,
these features are a common characteristic of
the interstellar medium (ISM) of the Milky Way
and nearby galaxies as well as distant galaxies
out to redshifts of $z\simgt4$
(e.g., see Riechers et al.\ 2014).
The ``UIE'' bands earned this name 
from the fact that the exact nature 
of their carriers remain unidentified,
although many candidate materials have been proposed.
Nevertheless, it is now generally accepted that these features
are characteristic of the stretching
and bending vibrations of some sorts of
aromatic hydrocarbon materials
(e.g., see Sellgren 2001, Peeters et al.\ 2002, 
Hudgins \& Allamandola 2005) 

L\'{e}ger \& Puget (1984) and
Allamandola et al.\ (1985, 1989)
attributed the UIE bands to gas-phase, free-flying
polycyclic aromatic hydrocarbon (PAH) molecules.
The PAH model attributes the UIE bands
to the vibrational modes of PAHs
with the 3.3$\mum$ feature assigned to
C--H stretching modes,
the 6.2$\mum$ and 7.7$\mum$ features to
C--C stretching modes,
the 8.6$\mum$ feature to
C--H in-plane bending modes,
and the 11.3$\mum$ feature to
C--H out-of-plane bending modes.
Alternatively, the UIE features have also
been attributed to amorphous solids with
a mixed aromatic/aliphatic composition
like hydrogenated amorphous carbon (HAC; Jones et al.\ 1990),
quenched carbonaceous composites (QCC; Sakata et al.\ 1990),
coal or kerogen (Papoular et al.\ 1989).\footnote{%
   The ``MAON'' model recently proposed
   by Kwok \& Zhang (2011, 2013) also falls
   in this category, where MAON stands for
   ``{\it mixed aromatic/aliphatic organic nanoparticle}''.
   }
As originally suggested by Duley \& Williams (1981),
all of these {\it solid} materials share
the basic molecular structure of PAHs
by containing arenes.
They also contain aliphatic C--H bonds as well as
other molecular structures often with other elements
besides C and H.

One way to distinguish the PAH hypothesis
from other models
is to compare in the UIE carriers $N_{\rm C,aliph}$, 
the number of carbon (C) atoms in aliphatic chains, 
with $N_{\rm C,arom}$, 
the number of C atoms in aromatic rings.
We define the {\it aliphatic fraction}
of the UIE carriers
as the fraction of C atoms in aliphatic chains.
For $N_{\rm C,aliph} \ll N_{\rm C,arom}$,
the aliphatic fraction is approximately
$N_{\rm C,aliph}/N_{\rm C,arom}$,
the ratio of the number of aliphatic C atoms
to that of aromatic C atoms.

%
%
%
%
%
Aliphatic hydrocarbons can be probed through
their C--H stretching vibrational band at 3.4$\mum$
(Pendleton \& Allamandola 2002).
%
%
%
In many interstellar and circumstellar environments
the 3.3$\mum$ emission feature is indeed often
accompanied by a weaker feature at 3.4$\mum$
(see Li \& Draine 2012 and references therein).
The 3.4$\mum$ emission feature itself is often 
also accompanied by several weak satellite features 
at 3.43, 3.46, 3.51, and 3.56$\mum$
(e.g., see Geballe et al.\ 1985, 
Jourdain de Muizon et al.\ 1986,
Joblin et al.\ 1996, 
Sloan et al.\ 1997, 2014,
Kaneda et al.\ 2014,
Mori et al.\ 2014,
Hammonds et al.\ 2015,
Ohsawa et al.\ 2016).
We define $I_{3.4}$ and $I_{3.3}$ 
as the observed intensities of the 3.4$\mum$
and 3.3$\mum$ emission features, respectively.
For those sources in which both the 3.4$\mum$ feature
and the 3.43, 3.46, 3.51, and 3.56$\mum$ satellite
features have been detected, $I_{3.4}$, 
the observed intensity of the 3.4$\mum$ emission feature,
also includes the contributions from the 3.43, 3.46, 3.51, 
and 3.56$\mum$ satellite features. 
We also define $A_{3.4}$ and $A_{3.3}$
as the band strengths
of the aliphatic and aromatic C--H bonds,
respectively, on a per unit C--H bond basis.
We note that $A_{3.4}$ includes
the contributions from all the aliphatic C--H
stretches in the $\simali$3.4--3.6$\mum$
wavelength range.
The ratio of the number of C atoms in aliphatic units
to that in aromatic rings in the UIE carriers can be 
derived from
$N_{\rm C,aliph}/N_{\rm C,arom}
\approx 0.3\times\,\left(I_{3.4}/I_{3.3}\right)
\times\,\left(A_{3.3}/A_{3.4}\right)$.\footnote{%
  The factor ``0.3'' arises from the conversion of 
  the number of C--H bonds ($N_{\rm C-H}$)
  to the number of C atoms ($N_{\rm C}$).
  For aliphatics, 
  we take $N_{\rm C,aliph} = 2.5 N_{\rm C-H,aliph}$
  (i.e., one aliphatic C atom corresponds 
   to 2.5 aliphatic C--H bonds)
  which is intermediate between methylene --CH$_2$ 
  and methyl --CH$_3$.
  For aromatics,
  we take $N_{\rm C,arom} = 0.75 N_{\rm C-H,arom}$
  (i.e., one aromatic C atom corresponds 
   to 0.75 aromatic C--H bond)
  which is intermediate between benzene C$_6$H$_6$
  and coronene C$_{24}$H$_{12}$.
  When converting $N_{\rm C-H,aliph}/N_{\rm C-H,arom}$
  to $N_{\rm C,aliph}/N_{\rm C,arom}$, a factor of
  $\left(0.75/2.5\right) = 0.3$ arises.
  }
As demonstrated by Li \& Draine (2012)
and Yang et al.\ (2013), one can place an upper limit
of $\simali$2\% on the aliphatic fraction of the emitters
of the UIE features by assigning the 3.4$\mum$ emission
{\it exclusively} to aliphatic C--H
(also see Rouill{\'e} et al.\ 2012,
Steglich et al.\ 2013).\footnote{%
  This is indeed an {\it upper limit}:
  the actual aliphatic fraction could be lower
  since the aliphatic C--H stretch is not 
  necessarily the sole contributor to 
  the 3.4$\mum$ emission feature.
  This feature could also be caused by 
  {\it anharmonicity} of 
  the aromatic C--H stretch (Barker et al.\ 1987)
  and ``{\it superhydrogenated}'' PAHs
  whose edges contain excess H atoms
  (Bernstein et al.\ 1996, Sandford et al.\ 2013).
  }
However, these studies were mostly based on
the $\Aratio$ ratios
derived from the {\it mono-methyl} derivatives of
small PAH molecules.
In reality, the PAH molecules in space
could harbor larger alkyl side chains
such as ethyl, propyl, and butyl.
Also, several alkyl side chains might 
be present in one PAH molecule.
To explore the effects of {\it larger}
alkyl side chains on $\Aratio$,
in this work we employ density functional theory
to compute the IR vibrational spectra of
several PAH molecules with a wide range of sidegroups
including ethyl, propyl, butyl,
and several unsaturated alkyl chains.
%
%
%
%
In \S\ref{sec:Method} we briefly describe
the computational methods.
The structures of the molecules
based on which we derive the band strengths
of the aromatic and aliphatic C--H stretches
are described in \S\ref{sec:Results}.
We compute their IR vibrational spectra
and report in \S\ref{sec:Results}
the calculated $\Aaro$ and $\Aali$
oscillator strengths and their implications
on the aliphatic fractions of the UIE carriers.
%
In view of the fact that interstellar PAHs
may have more than one alkyl side chain,
we also consider in \S\ref{sec:Results}
the situation that there are two methyl groups
attached to a PAH molecule,
using pyrene as an example.
We summarize our major results
in \S\ref{sec:summary}.

\section{Computational Methods
         \label{sec:Method}
         }
We use the Gaussian09 software (Frisch et al.\ 2009)
to calculate the IR vibrational spectra for a range of
aromatic molecules with a wide range of sidegroups
(see Figure~\ref{fig:LongSideChain}).
We employ the hybrid density functional
theoretical method (B3LYP)
at the {\rm 6-311+G$^{\ast\ast}$} level.
The standard scaling is applied to
the frequencies by employing
a scale factor of $\simali$0.9688 (Borowski 2012).
Yang et al.\ (2013) have carried out
computations for seven methylated PAH molecules
with the B3LYP method
in conjunction with a variety of basis sets.
These basis sets,
in order of increasing computational demand
and accuracy, are
   {\rm 6-31G$^{\ast}$},
   {\rm 6-31+G$^{\ast}$},
   {\rm 6-311+G$^{\ast}$},
   {\rm 6-311G$^{\ast\ast}$},
   {\rm 6-31+G$^{\ast\ast}$},
   {\rm 6-31++G$^{\ast\ast}$},
   {\rm 6-311+G$^{\ast\ast}$},
   {\rm 6-311++G$^{\ast\ast}$},
   {\rm 6-311+G(3df,3pd)}, and
   {\rm 6-311++G(3df,3pd)}.
The second-order M$\o$ller-Plesset
perturbation theory (MP2)
is widely considered to be more accurate than B3LYP
in computing band intensities
(see Cramer et al.\ 2004).
To examine the accuracy of the B3LYP method,
Yang et al.\ (2013) have performed calculations
with {\rm B3LYP} and {\rm MP2},
using the most extensive basis set 
{\rm 6-311++G(3df,3pd)} for both methods.
It is found that the results
computed from {\rm B3LYP/6-311++G(3df,3pd)}
closely agree with that from
{\rm MP2/6-311++G(3df,3pd)}.
On the other hand, 
Yang et al.\ (2013) have also found that the results
computed with the {\rm B3LYP} method
combined with
the four most sophisticated basis sets
   {\rm 6-311+G$^{\ast\ast}$},
   {\rm 6-311++G$^{\ast\ast}$},
   {\rm 6-311+G(3df,3pd)}, and
   {\rm 6-311++G(3df,3pd)}
essentially reach the convergence limit.
The {\rm B3LYP/6-311+G$^{\ast\ast}$} method
therefore presents an excellent compromise
between accuracy and computational demand.
Therefore, in this work we adopt
the {\rm B3LYP/6-311+G$^{\ast\ast}$} method.
%

\section{Results and Discussion\label{sec:Results}}
\begin{figure*}
\centerline{
\includegraphics[scale=0.4,clip]{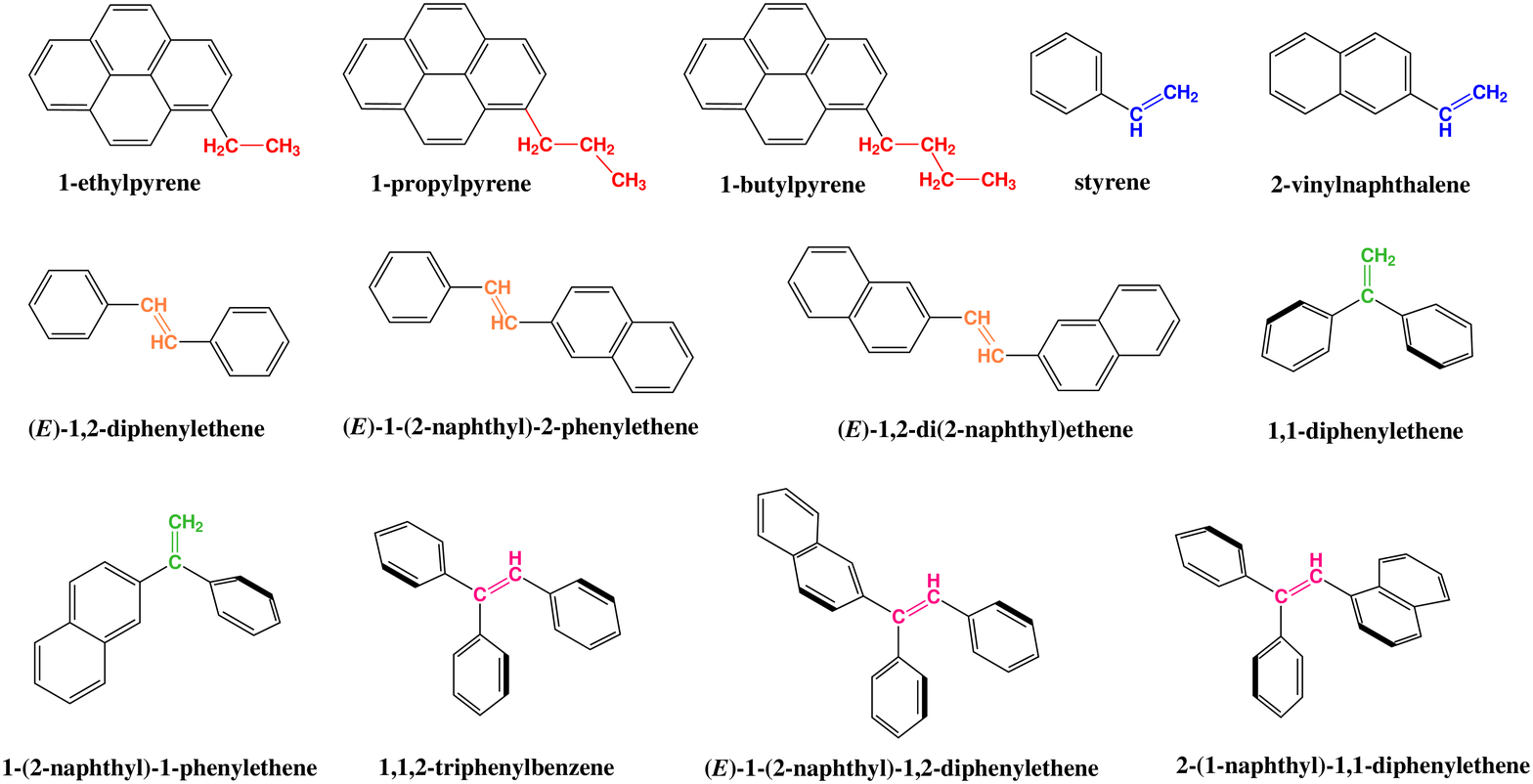}
}
\caption{\footnotesize
         \label{fig:LongSideChain}
         Effects of the nature of the alkyl group on
         the $\Aratio$ ratio were examined
         for 1-(n-alkyl)pyrene by consideration of
         ethyl (--CH$_2$--CH$_3$),
         propyl (--CH$_2$--CH$_2$--CH$_3$), and
         butyl (--CH$_2$--CH$_2$--CH$_2$--CH$_3$).
         Effects of olefinic C--H bonds on
         $\Aratio$ were examined for a number
         of vinyl-subsituted systems
         (--CH=CH$_2$)
         and for a number of systems with ethene spacers
         (--CH=CH--, C=CH$_2$, CH=C--H).
         }
\end{figure*}

\begin{figure}
\centerline
{
\includegraphics[width=15.6cm,angle=0]{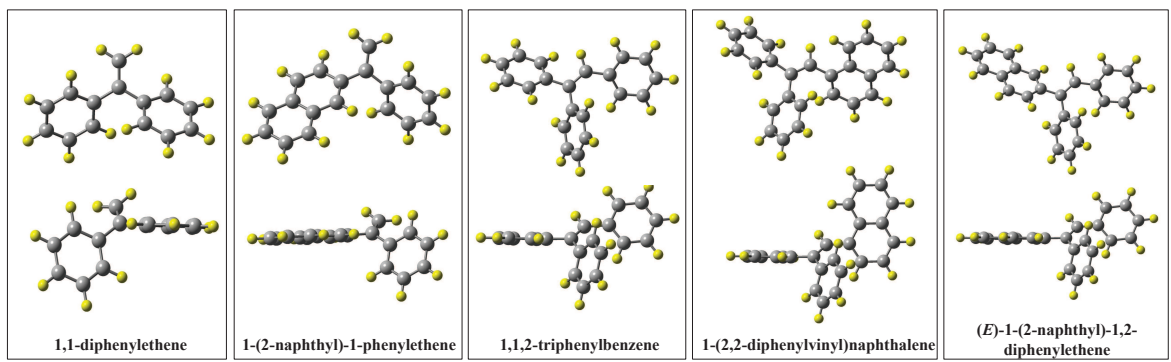}
}
\caption{\footnotesize
         \label{fig:OptimStruct}
         Optimized structures of phenyl- and
         naphthyl-substituted ethene.
         H atoms are shown in yellow and C atoms in grey.
         All structures are minima.
         }
\end{figure}

All of the molecules shown in
Figure~\ref{fig:LongSideChain}
are studied
at the {\rm B3LYP/6-311+G$^{\ast\ast}$} level.
They cover a wide range of sidegroups,
including ethyl (--CH$_2$--CH$_3$),
propyl (--CH$_2$--CH$_2$--CH$_3$),
butyl (--CH$_2$--CH$_2$--CH$_2$--CH$_3$),
and several unsaturated alkyl groups and spacers
(--CH=CH$_2$, --CH=CH--, C=CH$_2$, C=C--H).
Most of the structures shown in
Figure~\ref{fig:LongSideChain}
are unremarkable and planar or very close to planar.
Molecular models of the structures with significant
twists around the (alkene)C--C(arene) bonds are shown
in Figure~\ref{fig:OptimStruct} in two perspectives.
The computed total energies
and the thermochemical parameters
are summarized
in Table~\ref{tab:E_ThermPara_LongChain}
for the minima.\footnote{%
    A structure on the potential energy surface is
    a ``stationary structure''
    if the net inter-atomic forces
    on each atom is acceptably close to zero.
    A ``minimum'' is a stationary structure
    for which a small distortion along any
    internal coordinate increases the energy
    (all curvatures are positive).
    }

\begin{figure*}
\centerline{
\includegraphics[scale=0.48,clip]{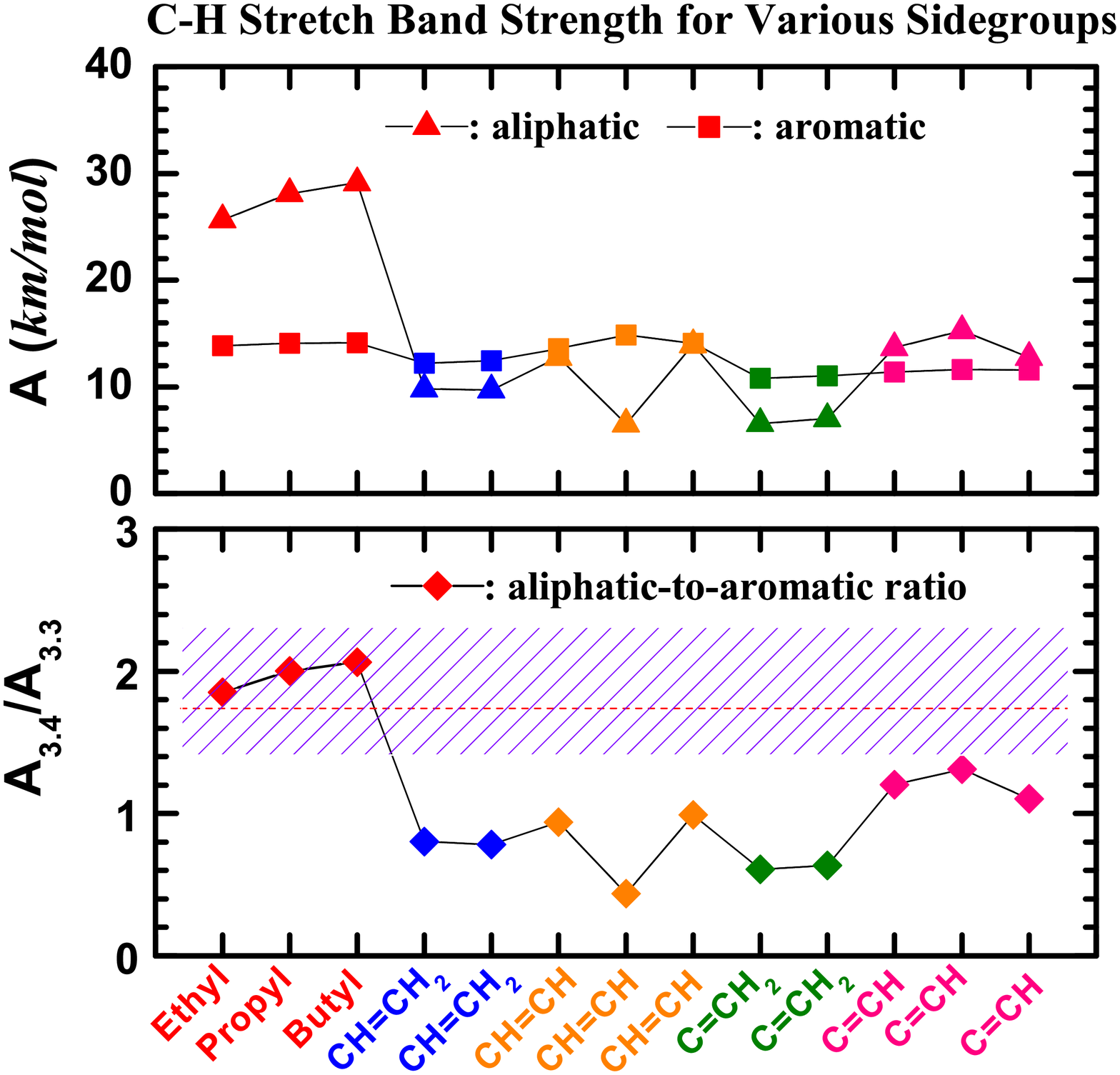}
}
\caption{\footnotesize
         Band-strengths as determined
         with the B3LYP/{\rm 6-311+G$^{\ast\ast}$} method
         for PAHs with sidegroups other than methyl:
         ethyl (--CH$_2$--CH$_3$),
         propyl (--CH$_2$--CH$_2$--CH$_3$),
         butyl (--CH$_2$--CH$_2$--CH$_2$--CH$_3$),
         and unsaturated alkyl chains
         (--CH=CH$_2$, --CH=CH--, C=CH$_2$, C=C--H).
         The shaded region shows the range of
         1.4\,$<$\,$\Aratio$\,$<$\,2.3
         derived for mono-methyl PAHs 
         which has a mean value
         of $\langle\Aratio\rangle\approx1.76$
         plotted here as a red dashed line
         (see Yang et al.\ 2013).
         \label{fig:A33A34_sidegroup_all}
         }
\end{figure*}

The vibrational frequencies and intensities for
the aromatic and the aliphatic C--H stretching
modes were computed.
The computed intensities are shown
in the top panel of Figure~\ref{fig:A33A34_sidegroup_all}.
We can see that the aliphatic C--H stretch
band strength varies within a wide range.
For ethyl, propyl and butyl,
the values ($\Aali$\,$\simali$25--30$\km\mol^{-1}$)
are generally consistent with methyl PAHs
for which the aliphatic C--H stretch
band strength has an average value
(per aliphatic C--H bond) of
$\langle \Aali\rangle \approx 23.68\km\mol^{-1}$,
with a standard deviation of
$\sigma(\Aali)\approx 2.48\km\mol^{-1}$
(see Yang et al.\ 2013).
In contrast, the aliphatic C--H stretch strengths
for the unsaturated alkyl chains
(--CH=CH$_2$, --CH=CH--, C=CH$_2$, C=C--H)
are much lower ($\Aali$\,$\simali$5--15$\km\mol^{-1}$).
On the other hand, the aromatic C--H stretch
band strength stays stable for all the groups,
$\Aaro$\,$\simali$10--15$\km\mol^{-1}$,
which is also consistent with methyl PAHs
for which the aromatic C--H stretch
band strength has an average value
(per aromatic C--H bond) of
$\langle \Aaro\rangle \approx 14.03\km\mol^{-1}$,
with a standard deviation of
$\sigma(\Aaro)\approx 0.89\km\mol^{-1}$
(see Yang et al.\ 2013).
Therefore, as shown in the bottom panel
of Figure~\ref{fig:A33A34_sidegroup_all},
we conclude that,
with $\langle \Aratio\rangle \approx 1.97\pm0.12$,
the $\Aratio$ ratios
for PAHs with ethyl, propyl and butyl groups
are close to that of methyl PAHs
for which $\langle \Aratio\rangle \approx 1.76\pm0.21$
(see Yang et al.\ 2013).
The $\Aratio$ ratios for PAHs with unsaturated
alkyl chains ($\langle \Aratio\rangle \approx 0.80\pm0.24$)
are lower by a factor of
$\simali$2 than that of methyl PAHs.
%
%

With the ratios of the intrinsic band strengths
$\Aratio$ computed,
we can estimate the ratio of the number of C atoms
in aliphatic units to that in aromatic rings
in the UIE carriers from
$N_{\rm C,aliph}/N_{\rm C,arom}
\approx 0.3\times\,\left(I_{3.4}/I_{3.3}\right)
\times\,\left(A_{3.3}/A_{3.4}\right)$.
Yang et al.\ (2013) have compiled and analyzed
the observed UIE spectra of 35 sources
reported in the literature
which display both the 3.3$\mum$
and 3.4$\mum$ C--H features.
These sources include PPNe, PNe, reflection nebulae, 
HII regions, and photodissociated regions (PDRs).
Yang et al.\ (2013) derived a median ratio of
$\langle\Iratio\rangle\approx 0.12$,
with the majority (31/35) of
these sources having $\Iratio < 0.25$.
With an average bond strength ratio
of $\Aali/\Aaro\approx 1.97$ derived
here for PAHs with ethyl, propyl and butyl sidegroups,
we obtain $N_{\rm C,aliph}/N_{\rm C,arom}\approx0.018$.
Even if we adopt $\Aali/\Aaro\approx 0.80$
which is derived for PAHs with unsaturated alkyl chains,
the aliphatic fraction is still only
$N_{\rm C,aliph}/N_{\rm C,arom}\approx0.045$.
This suggests that, in agreement of the earlier
findings of Li \& Draine (2012) and Yang et al.\ (2013),
the UIE emitters are predominantly aromatic.
An aliphatic component is indeed present,
as revealed by the 3.4$\mum$ C--H stretch,
but it is only a very minor
part of the UIE emitters.
%

%

%
More recently, Mori et al.\ (2014) obtained 
the $\simali$2.5--5.4$\mum$ emission spectra 
of 36 Galactic HII regions 
(or HII region-like objects) 
with the {\it Infrared Camera} (IRC)
on board {\it AKARI} (Onaka et al.\ 2007). 
They determined the 3.4\,$\mu$m-to-3.3\,$\mu$m
band ratios $\Iratio$ in a uniform manner 
for all 36 HII regions and derived an average ratio
of $\langle\Iratio\rangle\approx 0.25\pm0.10$
(T.~Onaka 2016, private communication),
which is $\simali$2.1 times 
that of Yang et al.\ (2013;
$\langle\Iratio\rangle\approx 0.12$).
Again, the 3.4$\mum$ feature intensity $\Iali$
includes the contributions from 
the weak satellite features at $\simali$3.4--3.5$\mum$.
For the UIE carriers in these HII regions, 
the aliphatic content will be higher:
$N_{\rm C,aliph}/N_{\rm C,arom}\approx0.038$
if one adopts the $\Aratio$ ratios derived 
for PAHs with ethyl, propyl and butyl groups
(i.e., $\Aratio\approx 1.97$),
or $N_{\rm C,aliph}/N_{\rm C,arom}\approx0.095$
if one adopts the $\Aratio$ ratios derived 
for PAHs with unsaturated alkyl chains
(i.e., $\Aali/\Aaro\approx 0.80$).
Apparently, in either case the UIR carriers
are still largely aromatic.
%

\begin{figure*}
\centerline{
\includegraphics[scale=0.7,clip]{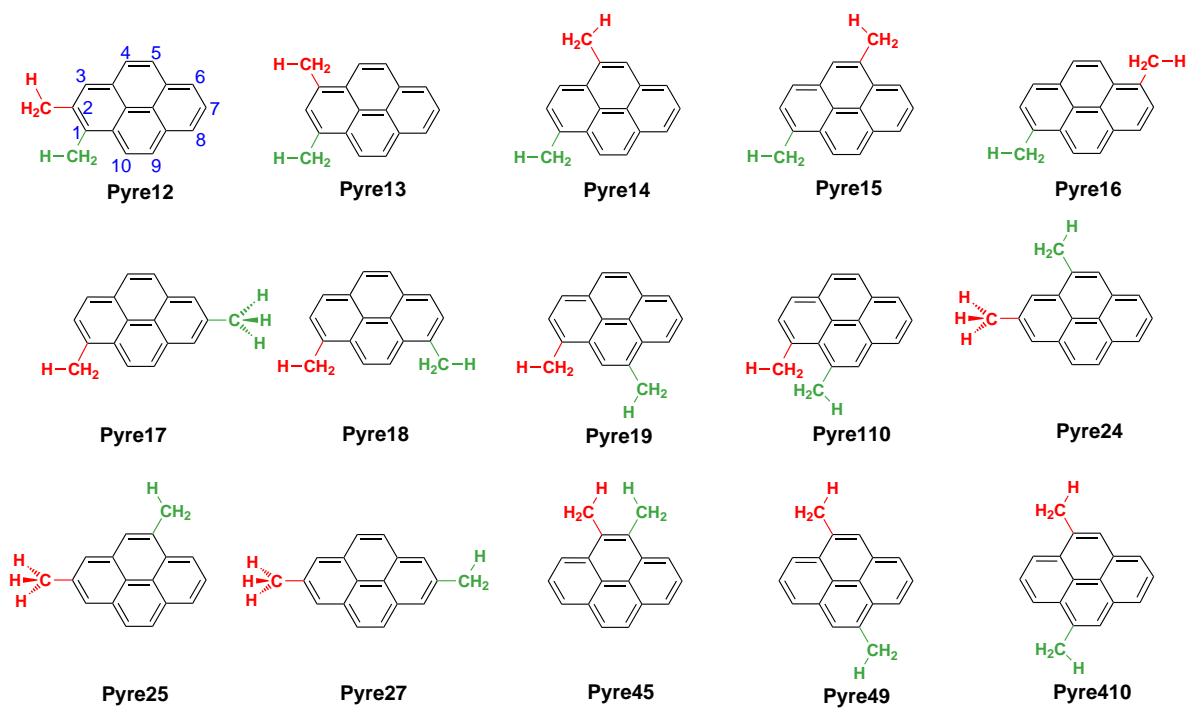}
}
\caption{\footnotesize
         Structures of all isomers of dimethylpyrene.
         The naming method follows the standard
         {\it International Union of Pure and Applied Chemistry}
         (IUPAC) numbering scheme.
         We use the first four letters of the molecule
         to refer to it and attach the position number
         of the location of the methyl group:
         ``Pyre'' stands for pyrene,
         and the digits specify the locations
         of the attached methyl groups
         (e.g., ``Pyre110'' means the two methyl
          groups are attached at positions 1 and 10).
          \label{fig:pyrene_dimethyl}
          }
\end{figure*}
\begin{figure*}
\centerline{
\includegraphics[scale=0.4,clip]{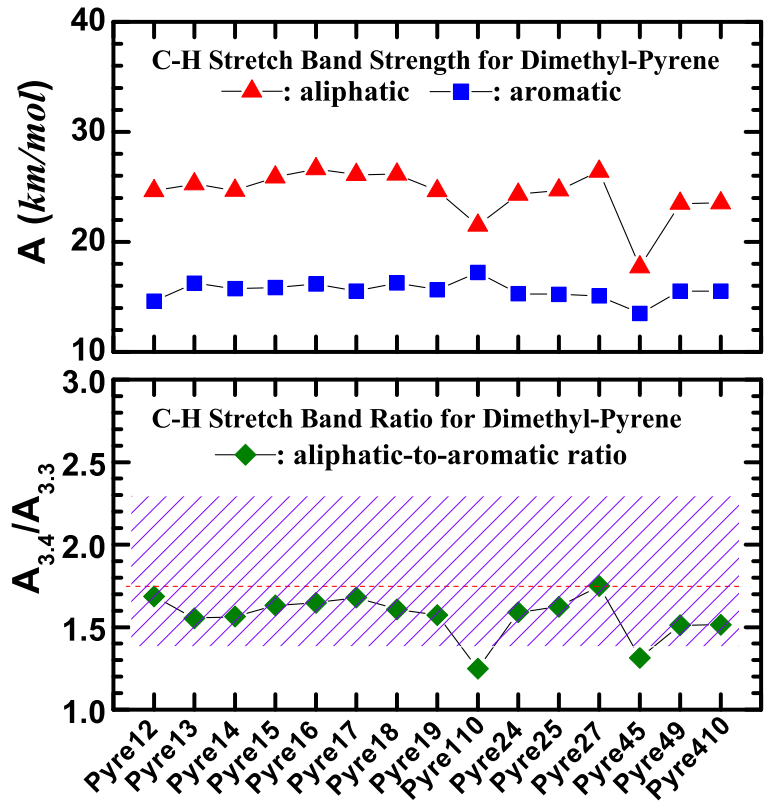}
}
\caption{\footnotesize
         \label{fig:A34A33_dimethyl_all}
         Band-strength as determined with
         the B3LYP/{\rm 6-311+G$^{\ast\ast}$} method
         for all the isomers of dimethyl pyrene.
         The shaded region shows the range of
         1.4\,$<$\,$\Aratio$\,$<$\,2.3
         derived for mono-methyl PAHs 
         which has a mean value
         of $\langle\Aratio\rangle\approx1.76$
         (red dashed line; see Yang et al.\ 2013).
         }
\end{figure*}

In view that several alkyl side chains
might be present in one PAH molecule,
we also consider the situation that there
are two methyl groups attached to a PAH molecule,
using pyrene as an example.
We consider all possible isomers of
dimethyl-substituted pyrene
(see Figure~\ref{fig:pyrene_dimethyl}).
For dimethyl pyrenes, as shown in the top panel
of Figure~\ref{fig:A34A33_dimethyl_all},
the aliphatic C--H stretch band strength
varies within $\Aali$\,$\simali$18--27$\km\mol^{-1}$,
while these values for the aromatic C--H stretch
are generally $\Aaro$\,$\simali$15$\km\mol^{-1}$.
The $\Aratio$ ratios,
as shown in the bottom panel
of Figure~\ref{fig:A34A33_dimethyl_all},
vary from $\simali$1.25 (Pyre110)
to $\simali$1.75 (Pyre27),
with an average ratio of
$\langle\Aratio\rangle\approx1.57\pm0.13$,
which is only $\simali$11\% lower
than the mean ratio of
$\langle\Aratio\rangle\approx1.76$
computed from methyl PAHs
(see Yang et al.\ 2013).
With $\langle\Iratio\rangle\approx 0.12$ 
(Yang et al.\ 2013)
or $\langle\Iratio\rangle\approx 0.25$ 
(Mori et al.\ 2014),
we derive the aliphatic fraction
of the UIE carriers
from $\langle\Aratio\rangle\approx1.57$
to be $N_{\rm C,aliph}/N_{\rm C,arom}\approx0.023$
or $N_{\rm C,aliph}/N_{\rm C,arom}\approx0.048$,
respectively.

We also find that the methyl groups are essentially 
independent of each other.
Noticeable effects on frequency and intensity only occur
when several alkyl groups are placed in direct proximity.
We note that for methyl PAHs the frequencies of
the aliphatic C--H stretch are always smaller than
$\simali$3000$\cm^{-1}$
and those for the aromatic C--H stretch
are larger than $\simali$3000$\cm^{-1}$.
The positions of the C--H stretches of
simple alkenes and dienes coincide with
the methyl signals of methyl-substituted PAHs.
However, for CH=CH$_2$ and C=CH$_2$,
one of the aliphatic C--H stretches
falls at $\simali$3120$\cm^{-1}$
(i.e, in the ``aromatic'' region).
For dimethyl pyrene Pyre45,
there is also one frequency of
the aliphatic C--H stretches that falls in
the ``aromatic'' region ($\simali$3070$\cm^{-1}$).
%

We note that although the 3.4$\mum$ 
{\it emission} feature is always much weaker than 
the 3.3$\mum$ {\it emission} feature 
in most of the UIE sources
(except several PPNe, 
see Geballe 1997, Tokunaga 1997),
the opposite is seen in {\it absorption}.
A broad {\it absorption} feature at 3.4$\mum$,
ubiquitously seen in the diffuse ISM of
the Milky Way and external galaxies
(see Sandford et al.\ 1991, 1995, 
Pendleton et al.\ 1994,
Pendleton \& Allamandola 2002),
is generally attributed to {\it solid} aliphatic 
hydrocarbon dust. It is much broader than 
the 3.4$\mum$ {\it emission} feature seen 
in the UIE sources (see Figure~1 in Li \& Draine 2012), 
supporting the {\it bulk, solid} nature
of the carriers of the 3.4$\mum$ absorption feature.
In some sources the 3.4$\mum$ absorption feature
is accompanied by a weaker 3.3$\mum$ {\it absorption}
feature (e.g., in the Galactic center source GCS 3, 
the 3.3$\mum$ absorption feature is weaker than 
the predominant 3.4$\mum$ feature by 
a factor of $\simali$35, see Chiar et al.\ 2000).
The 3.4$\mum$ {\it absorption} feature 
of the diffuse ISM consists of 
three substructures at $\simali$3.38, 3.42, 3.48$\mum$
(corresponding to $\simali$2955, 2925, 2870$\cm^{-1}$)
which respectively arises from 
the asymmetric C--H stretches of 
the --CH$_3$ and --CH$_{2}$-- 
functional groups 
and the symmetric C--H stretch 
of the --CH$_3$ groups
in saturated aliphatic hydrocarbon materials
(see Pendleton \& Allamandola 2002).
Some sources also exhibit a subfeature
at $\simali$3.51$\mum$ 
(corresponding to $\simali$2850$\cm^{-1}$),
attributed to the symmetric C--H stretch 
of the --CH$_{2}$-- groups.
Pendleton \& Allamandola (2002) derived
an abundance ratio of 
--CH$_{2}$--/--CH$_3$\,$\approx2.5$,
suggesting that the aliphatic hydrocarbon 
materials contain structures like
--CH$_2$--CH$_2$--CH$_3$ and
--CH$_2$--CH$_2$--CH$_2$--CH$_3$.
However, it is more complicated to infer
the possible structures of 
the aliphatic components
of the UIE carriers 
from the relative strengths 
of the 3.4$\mum$ emission feature and 
its associated satellite features
at 3.43, 3.46, 3.51 and 3.56$\mum$.
This is because the 3.4$\mum$ emission 
feature and its satellite features
could also be due to superhydrogenation
(Bernstein et al.\ 1996, Sandford et al.\ 2013)
and anharmonicity (Barker et al.\ 1987).
We stress that the carriers of 
the (stronger) 3.3 and (weaker) 3.4$\mum$
{\it emission} features 
cannot be the same carriers of
the (weaker) 3.3 and (stronger) 3.4$\mum$ 
{\it absorption} features.
While the former must be mostly aromatic,
nano-sized and subject to stochastic heating 
by single photons (see Sellgren et al.\ 1983, 
Sellgren 1984, Puget et al.\ 1995, Draine \& Li 2001),
the latter are most likely sub-$\mu$m-sized 
and attain equilibrium temperatures in the ISM.

While it is widely accepted that 
the 3.4$\mum$ {\it absorption} feature 
originates from aliphatic C--H stretch,
it remains debated whether the major 
carbon-containing interstellar dust component
is aliphatic or aromatic (see Furton et al.\ 1999, 
Greenberg \& Li 1999, Draine 2007).
%
%
Dartois et al.\ (2007) analyzed 
the $\simali$3.4--4.0$\mum$ spectrum
of IRAS~08572+3915,
a distant IR galaxy
at redshift $z\sim0.0583$,
obtained with the $L$-band spectrometer
of the {\it United Kingdom Infrared Telescope} (UKIRT).
Let $\tauaro\equiv\int \Delta\tau_{3.3}(\lambda)\,d\lambda$
be the optical depth ($\Delta\tau_{3.3}$)
of the 3.3$\mum$ aromatic C--H absorption feature   
integrated over wavelength, and
$\tauali\equiv\int \Delta\tau_{3.4}(\lambda)\,d\lambda$
be the optical depth ($\Delta\tau_{3.4}$)
of the 3.4$\mum$ aliphatic C--H absorption feature   
integrated over wavelength.
From the 3.3 and 3.4$\mum$ {\it absorption} features
of IRAS~08572+3915,
Dartois et al.\ (2007) determined 
$\tauali/\tauaro\approx33.3$ 
and further derived 
$N_{\rm H,aliph}/N_{\rm H,arom}\simgt12.5$
(which corresponds to 
$N_{\rm C,aliph}/N_{\rm C,arom}\simgt5.0$),
where $N_{\rm H,aliph}$ and $N_{\rm H,arom}$ 
are respectively the numbers of H atoms in
aliphatic chains and in aromatic rings.
Therefore, they argued
that {\it ``the bonding of hydrogen atoms in 
interstellar hydrogenated amorphous carbon 
is highly aliphatic''} (cf. Dartois et al.\ 2007). 
More recently,
Chiar et al.\ (2013) examined 
the $\simali$2.9--3.64$\mum$ spectrum
of the diffuse ISM along the line of sight
toward the Galactic center Quintuplet cluster
obtained with the UKIRT 
{\it Cooled Grating Spectrometer} (CGS\,4). 
Taking a set of intrinsic strengths
for the C--H stretches 
{\it larger} than that of Dartois et al.\ (2007)
but comparable to that of this work,\footnote{%
   The intrinsic strengths of 
   the symmetric C--H stretches of 
   the --CH$_2$-- and --CH$_3$ groups
   adopted by Chiar et al.\ (2013) 
   are respectively $\simali$7 
   and $\simali$12 times as high 
   as that of Dartois et al.\ (2007).
   The intrinsic strengths 
   of the asymmetric C--H stretches of 
   the --CH$_2$-- and --CH$_3$ groups
   as well as the aromatic C--H stretch
   adopted by Chiar et al.\ (2013) 
   are all about twice as high 
   as that of Dartois et al.\ (2007).
   }
they derived
$N_{\rm H,aliph}/N_{\rm H,arom}\approx1.19$
(which corresponds to 
$N_{\rm C,aliph}/N_{\rm C,arom}\approx0.48$)
from $\tauali/\tauaro\approx56.8$.

Nevertheless, taking into account  
the 6.2$\mum$ aromatic C--C stretching feature
observed both in the Galactic center Quintuplet
sightline and in IRAS~08572+3915,
Chiar et al.\ (2013) and Dartois et al.\ (2007) 
both concluded that the major carbon-containing
interstellar dust component is largely {\it aromatic}
(e.g., $N_{\rm C,aliph}/N_{\rm C,arom}\approx0.087$
for the Galactic center Quintuplet region,
see Chiar et al.\ 2013).
Chiar et al.\ (2013) suggested that 
interstellar hydrocarbon dust may consist of
a {\it large aromatic core} and 
a {\it thin aliphatic mantle}
(but also see Jones et al.\ [2013] who argued
for a {\it large aliphatic core} and 
a {\it thin aromatic mantle} 
for sub-$\mu$m-sized hydrocarbon dust).
If the carrier of the 3.4$\mum$ absorption feature
seen in the diffuse ISM is indeed just a {\it thin} layer
of aliphatic hydrocarbon material coated on a large
aromatic core (instead of a {\it thick} mantle coated on
an amorphous silicate core), 
the nondetection of the 3.4$\mum$ aliphatic 
C--H absorption polarization
along the line of sight where 
the 9.7$\mum$ Si--O silicate feature polarization 
has been detected (Chiar et al. 2006)
is no longer inconsistent with
the core-mantle model, 
provided that the mantle coated on 
the silicate core is not aliphatic but aromatic. 
If the silicate core is surrounded by 
an aliphatic hydrocarbon mantle, 
one would expect the 3.4$\mum$ absorption feature
to be polarized whereever the 9.7$\mum$ silicate 
feature is polarized
(see Li \& Greenberg 2002, Li et al.\ 2014). 

\section{Summary}\label{sec:summary}
The UIE carriers play an essential role in astrophysics
as an absorber of the UV starlight,
as an agent for photoelectrically
heating the interstellar gas,
and as a valid indicator of
the cosmic star-formation rates.
While the exact nature of the UIE carriers
remains unknown, the ratios of the observed intensities
of the 3.3$\mum$ aromatic C--H stretching emission feature
($\Iaro$) to that of the 3.4$\mum$ aliphatic C--H
emission feature ($\Iali$) could provide constraints
on the chemical structures of the UIE carriers,
i.e., are they mainly aromatic or largely aliphatic
with a mixed aromatic/aliphatic structure?
To this end, the knowledge of the intrinsic strengths
(per chemical bond)
of the 3.3$\mum$ aromatic C--H stretch ($\Aaro$)
and the 3.4$\mum$ aliphatic C--H stretch ($\Aali$)
is required. While the intrinsic strengths
of these C--H stretches were previously
derived almost exclusively from
the mono-methyl derivatives of small PAHs,
it is the purpose of this work to
derive $\Aratio$ based on extensive computations
of the vibrational frequencies and intensities of
a range of PAHs with side groups other than methyl
using the hybrid density function theory (B3LYP)
in conjunction with
the {\rm 6-311+G$^{\ast\ast}$} basis set.
The major results are:
\begin{enumerate}
\item A wide range of sidegroups
      (other than methyl)
      have been considered,
      including ethyl (--CH$_2$--CH$_3$),
      propyl (--CH$_2$--CH$_2$--CH$_3$),
      butyl (--CH$_2$--CH$_2$--CH$_2$--CH$_3$)
      and several unsaturated alkyl chains
      (--CH=CH$_2$, --CH=CH--, C=CH$_2$, C=C--H).
      With $\langle\Aratio\rangle\approx1.97$,
      the corresponding $\Aratio$ ratios are close
      to that of mono-methyl PAHs
      (for which $\langle\Aratio\rangle\approx1.76$),
      except PAHs with unsaturated alkyl chains
      (for which, with an average ratio of
       $\langle\Aratio\rangle\approx0.80$,
       the $\Aratio$ ratios could be lower
       by a factor of $\simali$2).
%
\item Dimethyl pyrene is studied in the context
      that several alkyl side chains
      might be present in one PAH molecule.
      The $\Aratio$ ratio averaged over
      all the isomers of dimethyl-substituted pyrene
      is $\simali$1.57, which is only $\simali$11\%
      lower than that of mono-methyl PAHs.
\item By attributing the 3.4$\mum$ feature
      exclusively to aliphatic C--H stretch
      (i.e., neglecting anharmonicity and
       superhydrogenation),
      we derive the fraction of C atoms
      in aliphatic form from $\Iratio$
      and $\Aali/\Aaro$,
      where $\Iratio$, the ratio of the power
      emitted from the 3.4$\mum$ feature to that from
      the 3.3$\mum$ feature, has a median ratio of
      $\langle\Iratio\rangle\approx 0.12$
      for 35 astronomical sources 
       (consisting of PPNe, PNe, reflection nebulae, 
        HII regions, and PDRs)
      which exhibit both the 3.3$\mum$ 
      and 3.4$\mum$ C--H features.
      With $\Aali/\Aaro\approx 1.97$
      for PAHs with ethyl, propyl and butyl sidegroups
      or $\Aali/\Aaro\approx 1.57$ for dimethyl pyrene,
      the derived aliphatic fractions of the UIE carriers
      are $\simali$2\%,
      in close agreement with that determined from
      methyl PAHs (for which $\Aali/\Aaro\approx 1.76$).
      Even with $\Aali/\Aaro\approx 0.80$
      derived for PAHs with unsaturated alkyl chains,
      the aliphatic fraction is only $\simali$4.5\%.
      We conclude that, in agreement of the earlier
      findings of Li \& Draine (2012) and Yang et al.\ (2013),
      the UIE emitters are predominantly aromatic
      and the aliphatic component is only a very minor
      part of the UIE emitters.
      This conclusion remains valid 
      for the UIE carriers 
      in the 36 Galactic HII regions
      studied by Mori et al.\ (2014)
      for which
      $\langle\Iratio\rangle\approx 0.25\pm0.10$.
\end{enumerate}

\acknowledgments{%
We thank Dr. Takashi Onaka 
and the anonymous referee
for very helpful suggestions.
AL and XJY are supported in part by
NSFC\,11473023,
NSF AST-1311804, NNX13AE63G,
and the University of Missouri Research Board.
RG is supported in part by NSF-PRISM grant
Mathematics and Life Sciences (0928053).
Computations were performed using the high-performance computer
resources of the University of Missouri Bioinformatics Consortium.
}



\begin{table*}
\footnotesize
\begin{center}
\caption[]{\footnotesize
           Computed Total Energies
           and Thermochemical Parameters
           for the Molecules Shown in
           Figure~\ref{fig:LongSideChain}
           at the {\rm B3LYP/6-311+G$^{\ast\ast}$} Level.
           }
\label{tab:E_ThermPara_LongChain}
\begin{tabular}{lccccccc}
\noalign{\smallskip} \hline \hline \noalign{\smallskip}
Compound	&	$\Etot$$^a$	
                &	VZPE$^b$	
                &	TE$^c$	
                &	$S$$^d$	
                &	$\nu_1$$^e$	
                &	$\nu_2$$^e$	
                &	$\mu$$^f$\\	
\noalign{\smallskip} \hline \noalign{\smallskip}
1-ethylpyrene	&	-694.563690 	&	164.68 	&	172.86 	&	111.66 	&	49.40 	&	99.48 	&	0.6960 	\\
1-propylpyrene	&	-733.887721 	&	182.33 	&	191.44 	&	119.15 	&	40.80 	&	89.70 	&	0.7423 	\\
1-butylpyrene	&	-773.212086 	&	200.09 	&	210.07 	&	126.68 	&	30.03 	&	72.90 	&	0.7933 	\\
styrene	&	-309.730792 	&	83.30 	&	87.59 	&	83.88 	&	22.17 	&	205.09 	&	0.2065 	\\
2-vinylnaphthalene	&	-463.409350 	&	112.66 	&	118.47 	&	94.45 	&	57.54 	&	132.81 	&	0.2748 	\\
(E)-1,2-diphenylethene	&	-540.846680 	&	134.22 	&	141.34 	&	109.43 	&	13.17 	&	57.63 	&	0.0000 	\\
(E)-1-(2-naphthyl)-2-phenylethene	&	-694.525418 	&	163.54 	&	172.24 	&	120.20 	&	22.42 	&	45.62 	&	0.0489 	\\
(E)-1,2-di(2-naphthyl)ethene	&	-848.202613 	&	192.60 	&	202.99 	&	135.13 	&	9.80 	&	31.38 	&	0.1256 	\\
1,1-diphenylethene	&	-540.839566 	&	134.16 	&	141.11 	&	105.16 	&	40.25 	&	60.61 	&	0.2926 	\\
1-(2-naphthyl)-1-phenylethene	&	-694.517262 	&	163.35 	&	171.95 	&	118.19 	&	32.50 	&	49.65 	&	0.3264 	\\
1,1,2-triphenylbenzene	&	-771.949489 	&	184.92 	&	194.84 	&	130.27 	&	30.52 	&	42.19 	&	0.3446 	\\
(E)-1-(2-naphthyl)-1,2-diphenylethene	&	-925.627411 	&	214.08 	&	225.67 	&	143.17 	&	28.83 	&	37.95 	&	0.3844 	\\
2-(1-naphthyl)-1,1-diphenylethene	&	-925.625007 	&	214.17 	&	225.73 	&	142.78 	&	23.86 	&	37.53 	&	0.3252 	\\
\hline
\noalign{\smallskip} \noalign{\smallskip}
\end{tabular}

\begin{description}
\item[$^{a}$] Total energies in atomic units.

\item[$^{b}$] Vibrational zero-point energies (VZPE)
                  in $\kcal\mol^{-1}$.

\item[$^{c}$] Thermal energies (TE) in $\kcal\mol^{-1}$.

\item[$^{d}$] Molecular entropies ($S$) in $\cals\mol^{-1}\K^{-1}$.

\item[$^{e}$] The lowest vibrational modes $\nu_1$ and $\nu_2$
                  in $\cm^{-1}$.

\item[$^{f}$] Dipole moment in Debye.

\end{description}
\end{center}
\end{table*}


\begin{table*}
\footnotesize
\begin{center}
\caption[]{\footnotesize
           Computed Total Energies
           and Thermochemical Parameters
           for All the Isomers of Dimethylpyrene
           Shown in Figure~\ref{fig:pyrene_dimethyl}
           at {\rm B3LYP/6-311+G$^{\ast\ast}$}.
           }
\label{tab:E_ThermPara_diPyrene}
\begin{tabular}{lccccccc}
\noalign{\smallskip} \hline \hline \noalign{\smallskip}
Compound	&	$\Etot$
                &	VZPE
                &	TE
                &	$S$
                &	$\nu_1$
                &	$\nu_2$
                &	$\mu$
\\ \noalign{\smallskip} \hline \noalign{\smallskip}
Pyre12	&	-694.404435 	&	165.41 	&	173.78 	&	112.18 	&	74.32 	&	89.80 	&	0.7983 	\\
Pyre13	&	-694.406226 	&	165.39 	&	173.77 	&	110.68 	&	87.00 	&	87.07 	&	0.5684 	\\
Pyre14	&	-694.407131 	&	165.47 	&	173.81 	&	111.63 	&	81.42 	&	105.91 	&	0.4788 	\\
Pyre15	&	-694.407075 	&	165.49 	&	173.82 	&	111.57 	&	86.33 	&	89.70 	&	0.0852 	\\
Pyre16	&	-694.406142 	&	165.42 	&	173.78 	&	110.45 	&	72.15 	&	111.02 	&	0.0001 	\\
Pyre17	&	-694.407577 	&	165.04 	&	173.61 	&	115.48 	&	33.38 	&	68.01 	&	0.3259 	\\
Pyre18	&	-694.406206 	&	165.38 	&	173.76 	&	112.02 	&	83.93 	&	94.45 	&	0.5851 	\\
Pyre19	&	-694.406998 	&	165.50 	&	173.82 	&	111.45 	&	73.06 	&	113.95 	&	0.5783 	\\
Pyre110	&	-694.396367 	&	165.75 	&	174.01 	&	112.73 	&	28.02 	&	100.75 	&	0.7623 	\\
Pyre24	&	-694.408443 	&	165.15 	&	173.67 	&	114.77 	&	36.84 	&	69.11 	&	0.7083 	\\
Pyre25	&	-694.408459 	&	165.18 	&	173.69 	&	114.75 	&	35.29 	&	73.52 	&	0.3881 	\\
Pyre27	&	-694.408977 	&	164.77 	&	173.51 	&	117.35 	&	29.34 	&	36.07 	&	0.0000 	\\
Pyre45	&	-694.400927 	&	165.72 	&	174.03 	&	110.48 	&	74.28 	&	76.98 	&	0.5827 	\\
Pyre49	&	-694.408023 	&	165.53 	&	173.83 	&	109.86 	&	73.49 	&	119.07 	&	0.0002 	\\
Pyre410	&	-694.407838 	&	165.56 	&	173.85 	&	111.16 	&	75.60 	&	121.61 	&	0.4010 	\\ \hline
\noalign{\smallskip} \noalign{\smallskip}
\end{tabular}
\end{center}
\end{table*}

\end{document}